\documentclass[letterpaper]{article}

\usepackage{arxiv}

\usepackage{amsmath}
\usepackage{amsfonts}
\usepackage{amssymb}

\usepackage{algorithm}
\usepackage{algpseudocode}

\usepackage[utf8]{inputenc} 
\usepackage[T1]{fontenc}    
\usepackage[hidelinks]{hyperref}
\usepackage[nameinlink,capitalize]{cleveref}
\usepackage{url}            
\usepackage{booktabs}       

\usepackage{nicefrac}       
\usepackage{microtype}      
\usepackage{graphicx}
\usepackage{caption}
\usepackage[numbers]{natbib}
\usepackage{doi}

\usepackage[section]{placeins} 

\usepackage[dvipsnames]{xcolor}
\usepackage{makecell}
\usepackage[version=4]{mhchem}
\usepackage[per-mode=symbol,round-precision=3,scientific-notation=false,range-phrase = \text{--}]{siunitx}

\usepackage{wrapfig}

\usepackage{framed} 
\usepackage{multicol} 
\usepackage{nomencl} 
\usepackage{ifthen}
\renewcommand\nomgroup[1]{%
  \ifthenelse{\equal{#1}{A}}{%
    \item[\textbf{Acronyms}]}{
  \ifthenelse{\equal{#1}{R}}{
    \item[\textbf{Roman Symbols}]}{
  \ifthenelse{\equal{#1}{G}}{%
    \item[\textbf{Greek Symbols}]}{
  \ifthenelse{\equal{#1}{S}}{%
    \item[\textbf{Superscripts}]}{
  \ifthenelse{\equal{#1}{U}}{%
    \item[\textbf{Subscripts}]}{
  \ifthenelse{\equal{#1}{X}}{%
    \item[\textbf{Other Symbols}]}{
  {}}}}}}}}
\renewcommand*{\nompreamble}{\markboth{\nomname}{\nomname}}

\makenomenclature
\renewcommand*\nompreamble{\begin{multicols}{2}}
\renewcommand*\nompostamble{\end{multicols}}

\newcommand\T{\rule{0pt}{2.6ex}}
\newcommand\B{\rule[-1.2ex]{0pt}{0pt}}

\usepackage{xspace}
\newcommand*{\eg}{e.g.,\@\xspace}
\newcommand*{\ie}{i.e.,\@\xspace}

\newcommand*{\vs}{vs.\@\xspace}

\makeatletter
\newcommand*{\etc}{%
    \@ifnextchar{.}%
        {etc}%
        {etc.\@\xspace}%
}
\makeatother

\DeclareUnicodeCharacter{025B}{\ensuremath{\varepsilon}}

\graphicspath{{figures/}}

\usepackage{listings}

\definecolor{bkgd}{RGB}{240,242,246}
\definecolor{ceruleanblue}{rgb}{0.16, 0.32, 0.75}
\definecolor{orange-red}{rgb}{1.0, 0.27, 0.0}
\definecolor{anotherblue}{RGB}{37,92,243}
\definecolor{blackblue}{RGB}{46,60,85}
\definecolor{goldyellow}{RGB}{199,146,12}
\lstdefinestyle{altstyle2}{
    backgroundcolor=\color{bkgd},
    basicstyle=\ttfamily\footnotesize\color{blackblue},
    breakatwhitespace=false,
    breaklines=true,
    captionpos=b,
    commentstyle=\color{goldyellow},
    keepspaces=true,
    keywordstyle=\color{orange-red},
    language=Python,
    numbersep=5pt,
    numberstyle=\tiny\color{ceruleanblue},
    showspaces=false,
    showstringspaces=false,
    showtabs=false,
    stringstyle=\color{anotherblue},
    tabsize=2,
    numbers=left
}
\usepackage[scaled=0.86]{helvet}

\crefname{lstlisting}{listing}{listings}
\Crefname{lstlisting}{Listing}{Listings}

\usepackage{pdflscape}

\DeclareMathOperator*{\argmin}{\arg\!\min}
\algnewcommand{\Inputs}[1]{%
  \State \textbf{Inputs:}
  \Statex \hspace*{\algorithmicindent}\parbox[t]{.8\linewidth}{\raggedright #1}
}
\algnewcommand{\Initialize}[1]{%
  \State \textbf{Initialize:}
  \Statex \hspace*{\algorithmicindent}\parbox[t]{.8\linewidth}{\raggedright #1}
}

\newcommand*{\qrx}{\dot{Q}_\mathrm{RX}}
\newcommand*{\qrxref}{\dot{Q}_\mathrm{RX,Ref}}
\newcommand*{\vqrx}{v_{\dot{Q}_\mathrm{RX}}}
\newcommand*{\vqrxmax}{v_{\dot{Q}_\mathrm{RX}}^\mathrm{max}}
\newcommand*{\qsg}{\dot{Q}_\mathrm{SG}}
\newcommand*{\qhx}{\dot{Q}_\mathrm{HX}}
\newcommand*{\rhoext}{\rho_\mathrm{ext}}
\newcommand*{\rhoc}{\rho_\mathrm{c}}
\newcommand*{\rhom}{\rho_\mathrm{m}}
\newcommand*{\rhosig}{\rho_\Sigma}
\newcommand*{\Tco}{T_\mathrm{c,out}}
\newcommand*{\Tci}{T_\mathrm{c,in}}
\newcommand*{\Pco}{P_\mathrm{c,out}}
\newcommand*{\Pci}{P_\mathrm{c,in}}
\newcommand*{\Tso}{T_\mathrm{s,out}}
\newcommand*{\Tsomax}{T_\mathrm{s,out}^\mathrm{max}}
\newcommand*{\Tsi}{T_\mathrm{s,in}}
\newcommand*{\Tsimin}{T_\mathrm{s,in}^\mathrm{min}}
\newcommand*{\mfp}{\dot{m}_\mathrm{p}}
\newcommand*{\mfs}{\dot{m}_\mathrm{s}}
\newcommand*{\dpp}{\Delta P_\mathrm{p}}
\newcommand*{\dps}{\Delta P_\mathrm{s}}

\newcommand*{\bA}{\mathbf{A}}
\newcommand*{\bB}{\mathbf{B}}

\newcommand*{\bx}{\mathbf{x}}

\definecolor{textred}{RGB}{170,0,0}

\usepackage{eqparbox}

\usepackage{tikz}
\usetikzlibrary{shapes.geometric, arrows}

\title{Design of a Supervisory Control System for Autonomous Operation of Advanced Reactors}

\author{ 
	Akshay J.~Dave \\
	Nuclear Science and Engineering\\
	Argonne National Laboratory\\
	Lemont, IL 60439\\
	\texttt{ajd@anl.gov} \\
	\And
	Taeseung Lee\\
	Nuclear Science and Engineering\\
	Argonne National Laboratory\\
	Lemont, IL 60439\\
	\And
	Roberto Ponciroli\\
	Nuclear Science and Engineering\\
	Argonne National Laboratory\\
	Lemont, IL 60439\\
	\And
	Richard B.~Vilim \\
	Nuclear Science and Engineering\\
	Argonne National Laboratory\\
	Lemont, IL 60439\\
}

\date{}

\renewcommand{\headeright}{ }
\renewcommand{\undertitle}{ }
\renewcommand{\shorttitle}{Design of a Supervisory Control System for Autonomous Operation of Advanced Reactors}

\begin{document}
\maketitle

\begin{abstract}

Advanced reactors to be deployed in the coming decades will face deregulated energy markets, and may adopt flexible operation to boost profitability.
To aid in the transition from baseload to flexible operation paradigm, autonomous operation is sought.
This work focuses on the control aspect of autonomous operation.
Specifically, a hierarchical control system is designed to support constraint enforcement during routine operational transients.
Within the system, data-driven modeling, physics-based state observation, and classical control algorithms are integrated to provide an adaptable and robust solution. 
A \SI{320}{\mega\watt} Fluoride-cooled High-temperature Pebble-bed Reactor is the design basis for demonstrating the proposed control system.

\vspace{1.0em}
The hierarchical control system consists of a supervisory layer and low-level layer.
The supervisory layer receives \textit{requests} to change the system's operating conditions (\eg the current reactor power to meet a load-follow), and accepts or rejects them based on constraints that have been assigned.
Constraints are issued to keep the plant within an optimal operating region.
The low-level layer interfaces with the actuators of the system to fulfill requested changes, while maintaining tracking and regulation duties.
To accept requests at the supervisory layer, the Reference Governor algorithm was adopted.
To model the dynamics of the reactor, a system identification algorithm, Dynamic Mode Decomposition, was utilized.
To estimate the evolution of process variables that cannot be directly measured (\eg the propagation of delayed neutron precursors), the Unscented Kalman Filter, incorporating a nonlinear model of nuclear dynamics, was adopted.
The composition of these algorithms led to a numerical demonstration of constraint enforcement during a 40\% power drop transient (at a rate of \SI{5}{\percent\per\minute}).
Uncontrolled secondary-side temperatures were successfully constrained.
Adaptability of the proposed system was demonstrated by modifying the constraint values, and enforcing them during the transient.
Robustness was also demonstrated by enforcing constraints under noisy environments.

\end{abstract}


\cleardoublepage
\section{Introduction}

Advanced reactors deployed in the coming decades will be operated in a grid with a larger portion of intermittent energy sources.
In the United States, renewable energy is projected to compose 20\% of the energy production mix by 2050 \citep{center2022annual}, a significant rise from 5\% in 2010, and 10\% currently.
The production from nuclear power plants (NPP) is expected to stay roughly flat or decline.
Besides the well-known high upfront capital costs, a reason of this trend might stem from the profitability issues that currently operated Light Water Reactors have been facing in the last years.
One late 2018 study found that more than one-third of units in the US is unprofitable or scheduled to close \citep{clemmer2018}.
It has been projected that flexible operation, as opposed to the current baseload operation paradigm, in a grid with significant intermittent sources, will reduce operating costs of NPPs and make nuclear energy a more competitive source \citep{jenkins2018benefits}.
When operated in flexible mode, the plant is expected to routinely undergo power transients to meet load demand fluctuations.
During these transients, the thermal-hydraulic and neutronic conditions of the NPP will change and require operators to adapt and carefully steer the plant.
To relieve human operators from this burden, autonomous operation of NPPs is sought.

The safe and reliable operation of an NPP requires effective teamwork, \ie the collaboration between the operators in the Main Control Room and the on-site technicians is the best way to produce power with fewer errors, events, and improved performance.
Similarly, implementation of the “autonomous operation” paradigm requires a tight integration of modules performing control, diagnostics and decision-making tasks.
The diagnostics module is tasked with processing sensor data and inferring the health and the performance of components, sensors and actuators.
The decision-making module is tasked with assigning high-level reference setpoints for the plant, \eg changing the current reactor power to meet a demand variation. 
The control module is tasked with steering the plant, coordinating control actions for all major actuators to address the setpoints that have been assigned, while meeting the constraints that have been assigned by the diagnostics module.
This work is focused on the design of a supervisory control system addressing the objectives of the control module.

\begin{wrapfigure}{l}{0.33\linewidth}
    \includegraphics[width=0.99\linewidth]{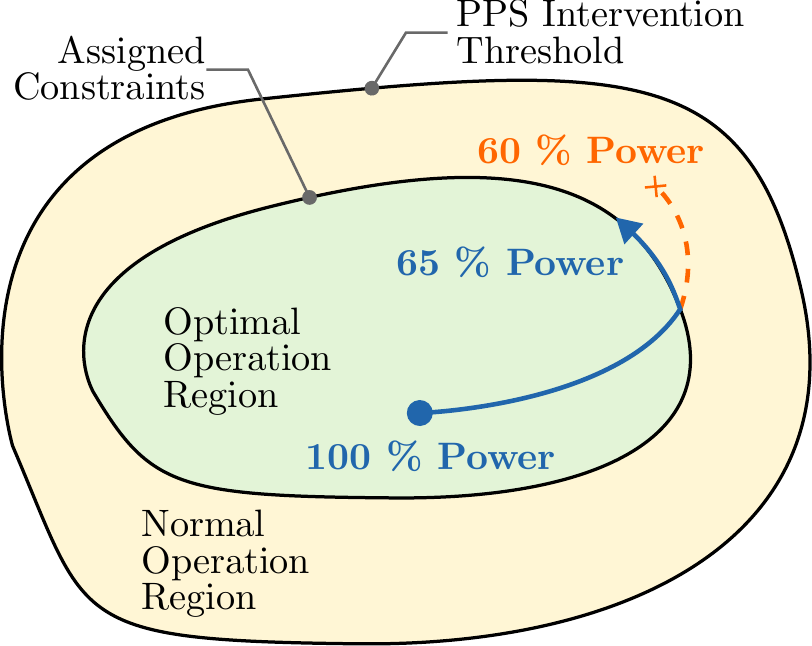}
    \caption{
    Graphical abstract of the application of a supervisory controller during a load-follow transient. 
    The supervisory controller steers the plant within the optimal operating region to meet high-level demands, while enforcing assigned constraints.
    }
    \label{fig:moas-abs}
\end{wrapfigure}
The idea of adopting a supervisory control algorithm for the operation of NPPs was originally proposed in the late 70s.
One of the first attempts was the hierarchical control system architecture designed for coordinating the actuators during load-follow transients in Clinch River Breeder Reactor (between 40 \% and 100 \% power) \citep{demore1977plant}. 
In the following years, other designs were proposed \citep{White1989,otaduy1991distributed,cetiner2016supervisory}.
The control system presented in this work also consists of a two-tier hierarchy: the supervisory layer and the low-level layer.
The major task of the former consists in accepting or rejecting changes in high-level setpoints (\cref{fig:moas-abs}), \eg a reactor power variation during a load-follow transient.
The acceptance or rejection is based on a quantitative trade-off of addressing requested changes to the system, while meeting the imposed constraints.
The output of the supervisory layer is sent to the low-level layer.
The low-level layer is constituted by controllers that operate in tracking or regulation mode.
Tracking controllers always accept changes in setpoints from the supervisory layer, \eg a proportional-integral-derivative (PID) controller to adjust reactor power via reactivity insertion.
Regulation controllers maintain setpoints, \eg PID control of a pump to maintain core inlet temperature.
NPPs are under-actuated systems, \ie the available actuators are fewer than the process variables to be controlled.
Thus, there may be process variables that cannot be controlled directly, \eg temperatures in the secondary-side, that would deviate from their full-power values. 
To reduce thermomechanical stresses and limit wear \& tear, excessive variations of these process variables should be avoided. 
An optimal operation region must be provided to the supervisory controller.
It quantitatively defines the bounds by which the supervisory controller can accept or reject changes to the high-level setpoints, such that constraints are not violated.
In \cref{fig:moas-abs}, a decrease from 100 \% to 60 \% load is requested.
However, due to an assigned constraint(s), power is not decreased below 65 \%.
The supervisory controller intervenes to avoid violating a constraint imposed by the operators, and the plant remains in the optimal operation region.
The intent of the supervisory layer is not to be a replacement for safety related interventions (conditions where the plant protection system (PPS) will trip the plant, \ie outside of the yellow region in \cref{fig:moas-abs}), but rather optimizing dynamics within the normal operation region.

The \textbf{objective} of this work is to design and demonstrate a supervisory control system that allows constraint enforcement during the load-follow transient for an advanced NPP.
In \cref{sec:gfhr}, performance of an advanced NPP under classic control is presented.
To address the shortcomings of classic control, a supervisory control system is designed.
To develop the constraint enforcement capability, two major components are needed, \ie (1) a forward model to predict the system response given the performed control actions, and (2) a method to estimate the admissible region given a set of constraints.
The methods that form the supervisory control system are presented in \cref{sec:methods}.
In \cref{sec:results}, performance of the plant under supervisory control is presented.
Lastly, concluding remarks and future work are described in \cref{sec:conclusions}.

\section{Classic Control of an Advanced Reactor}\label{sec:gfhr}

\begin{figure}
        \begin{minipage}[t]{0.7\linewidth}
        \vspace{0pt}
        \centering
        \includegraphics[width=\linewidth]{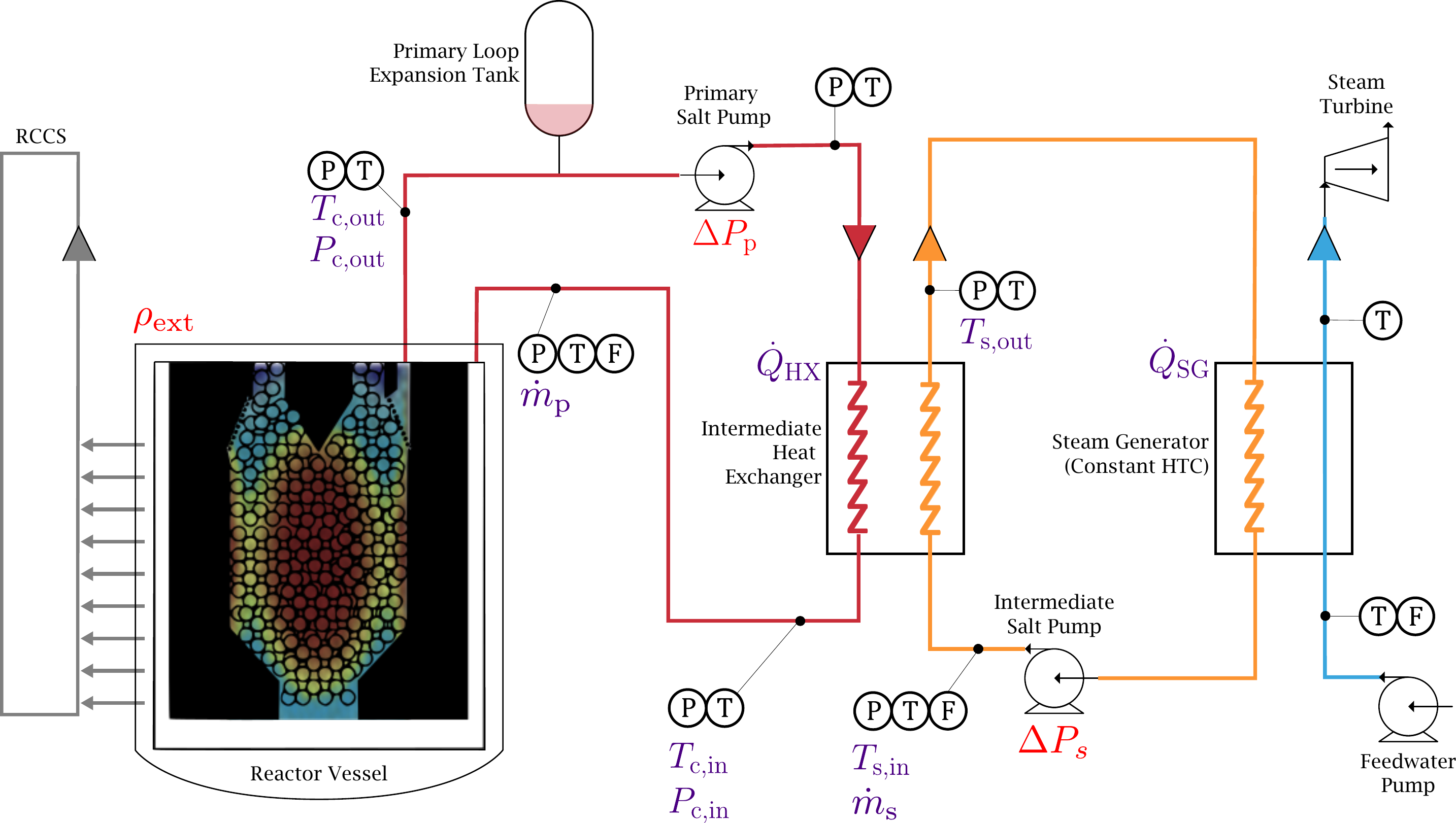}
        \end{minipage}%
        \begin{minipage}[t]{0.3\linewidth}
        \vspace{12pt}
        \centering
         \begin{tabular}{lc}
             \hline 
             \multicolumn{2}{c}{Full Power Values} \T\B\\
             \hline
             \multicolumn{2}{l}{Primary} \T\\
              \hspace{6pt}$\qrx$ [MW] & 320 \\
              \hspace{6pt}$\qhx$ [MW] & 313 \\
              \hspace{6pt}$\mfp$ [kg/s] & 1320 \\
              \hspace{6pt}$\Tci$ [\SI{}{\celsius}] & 547 \\
              \hspace{6pt}$\Tco$ [\SI{}{\celsius}] & 645 \\
              \hspace{6pt}$\Pci$ [kPa] & 1151 \\
              \hspace{6pt}$\Pco$ [kPa] & 156 \\
            \multicolumn{2}{l}{Secondary} \T\\
              \hspace{6pt}$\qsg$ [MW] & 313 \\
              \hspace{6pt}$\mfs$ [kg/s] & 5295 \\
              \hspace{6pt}$\Tsi$ [\SI{}{\celsius}] & 430 \\
              \hspace{6pt}$\Tso$ [\SI{}{\celsius}] & 469 \\
         \end{tabular}
        \end{minipage}
    \caption{
        Overview of the gFHR SAM model.
        Left: Piping and instrumentation diagram for the gFHR model.
        States that are required by the supervisory algorithm are identified in \textcolor[HTML]{4b0082}{purple} and actuators that are used to control the plant are identified in \textcolor[HTML]{ff0000}{red}.
        The control strategy and selection of states is discussed in \cref{subsec:strategy,subsec:statesel}, respectively.
        Right: A list of full power values corresponding to the labels in the left-hand side diagram.
        }
    \label{fig:gfhr-pid}
\end{figure}

To demonstrate the advantages of adding a supervisory control layer to traditional architectures, the performance of an advanced reactor under the classical control method is presented. 
In this work, the Fluoride-cooled High-Temperature pebble-bed Reactor (FHR) design was selected (it is also the design basis for Kairos Power \cite{blandford2020kairos}).
Significant efforts at Argonne National Laboratory \citep{hu2020development,o2021sam} have been directed at developing a generic FHR (gFHR) model to foster the design and safety analyses of the FHR concept. 
To simulate the plant response during operational transients, a high-fidelity model was developed by using the System Analysis Module (SAM) \cite{Hu2017} system code.
SAM is developed to provide best-estimate system-level models for Sodium-cooled Fast Reactors, Molten Salt Reactors, Fluoride-cooled High-temperature Reactors, and Lead-cooled Fast Reactors.
The gFHR is a hypothetical \SI{320}{\mega\watt}th reactor with a flibe primary loop, and a solar salt intermediate loop.
The reactor core and plant design choices are detailed in \citep[\S 3]{o2021sam}.
A Reactor Cavity Cooling System (RCCS) is incorporated in the model.
The RCCS system has tall chimneys exposed to the environment (air) to aid buoyancy driven passive cooling of the reactor vessel in case of an accident.
In \cref{fig:gfhr-pid}, the piping and instrumentation diagram (P\&ID) for the gFHR model is presented.
To mimic the presence of an energy conversion cycle, a pipe with a fixed heat transfer coefficient is present between the intermediate heat exchanger and the intermediate salt pump.

\subsection{Control Strategy for Load-follow} \label{subsec:strategy}

Advanced NPPs in future electrical grids will be expected to routinely perform load-follow transients on a daily, even hourly basis \citep{international2018non}.
Undergoing cyclical changes in baseload power, on the order of tens to hundreds of Mega Watts, would result in thermal stresses on components throughout the primary circuit.
To demonstrate the feasibility of flexible operation paradigm and reduce O\&M costs, these cyclical stresses must be minimized. 
At the extremes, we want to mitigate the possibility of thermal-fatigue induced failure leading to outages or accidents \cite{metzner2005european}.

In this work, we have designed a control strategy that mitigates thermal cycling during load-follow operation.
The quantitative objective of the control strategy is to regulate core inlet and outlet temperatures.
In order to manipulate the reactor power, external reactivity insertion will be used to address requests in load increase/decrease.
Pumps on both primary and secondary sides will then be used to manage heat transfer rates across the primary heat exchanger.
The pairings between Control Variables (CVs) and Manipulated Variables (MVs) were chosen to maximize the effects of control actions and to minimize cross-talk.
The pairings are listed below and tabulated in \cref{tab:vars}.
\begin{enumerate}
    \item The reactor power ($\qrx$) is controlled by external reactivity insertion ($\rhoext$). 
    \item The core outlet temperature ($\Tco$) is controlled by the primary pump head ($\dpp$).
    \item The core inlet temperature ($\Tci$) is controlled by the solar salt pump head ($\dps$).
\end{enumerate}
Once the appropriate pairing options are selected, three PID controllers were designed, \ie two PID controllers regulate the core outlet and inlet temperatures, and another one tracks the reactor power.
PIDs were manually tuned, first by following the Ziegler-Nichols \citep{hang1991refinements} method, followed by fine-tuning via a grid search of the proportional and integral gains.
The derivative term is nullified for all controllers.

\begin{table}%
    \centering
    \caption{%
    Pairing of control variables (CV) to manipulated variables (MV) to address objectives of the control strategy.
    }
    \begin{tabular}{c}%
        \begin{tabular}[t]{ll}
            \hline\T\B
            CVs & MVs \\
            \hline\T\B
            $\rhoext$ & $\qrx$ \\
            $\dpp$ & $\Tco$ \\
            $\dps$ & $\Tci$ \\
        \end{tabular}
    \end{tabular}
    \label{tab:vars}
\end{table}

To apply the proposed supervisory control system, constraints on both input and output variables need to be defined.
Any number of constraints may be applied, and they may be either constant or time-dependent.
The latter is a powerful capability in the perspective of developing an Autonomous Operation architecture.
If the control system is coupled to a diagnostics/prognostics system, constraints may be updated due to degradation of component performance (\eg leaks in a pump or fouling of a heat exchanger).
The proposed control strategy foresees three major actuators to govern the dynamics of the primary loop.
However, the secondary side of the plant is not directly controlled.
In \cref{fig:gfhr_sch}, the bottom row shows that the secondary inlet and outlet temperatures ($\Tsi$ and $\Tso$), reach new equilibrium values after the transient is completed.
Experimental studies show that corrosion of metals in contact with salt is strongly dependent on temperature \citep{sabharwall2014advanced,zheng2015corrosion}, and that optimal operating temperatures to mitigate weight loss exist.
Thus, issuing constraints to avoid excessive temperature variations would be a practical application of the supervisory control system. 
Two constraints are adopted on the secondary side: a minimum temperature constraint on the heat exchanger inlet and a maximum temperature constraint on the heat exchanger outlet.
The results from enforcing these constraints using the supervisory layer will be presented in \cref{subsec:rg_res}.

\subsection{Plant Performance during Load Follow}

In this section, the performance of the gFHR during load-follow transients \textit{without a supervisory layer} is discussed.
Throughout all cases studied, the rate at which power drops is \SI{5}{\percent\per\minute}.
As discussed in the previous subsection, three PID controllers are implemented to adjust the reactor power while maintaining the core outlet and inlet temperatures close to the full power values to preserve the thermal conditions in the reactor core.

\begin{figure}
    \centering
    \includegraphics[width=\linewidth]{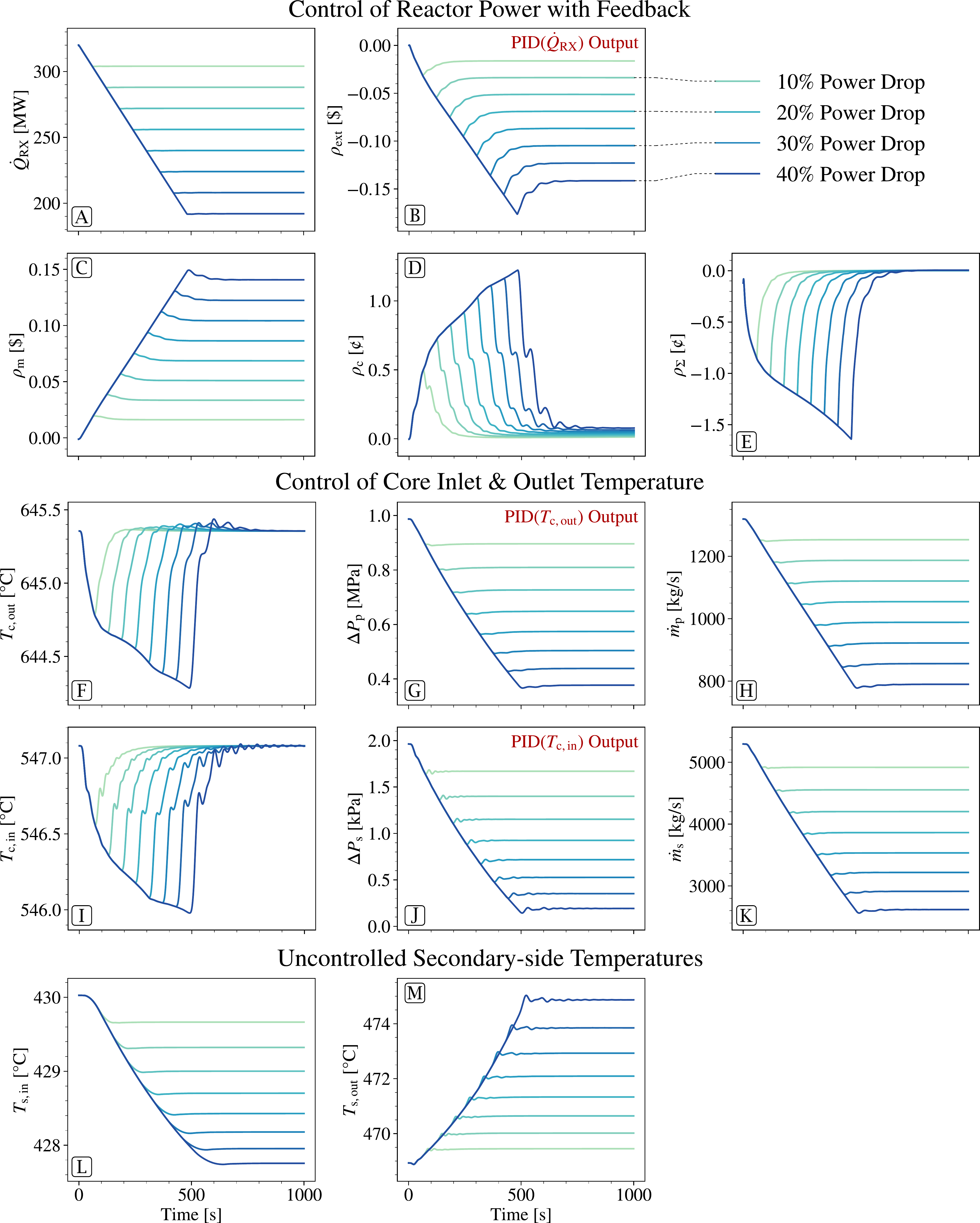}
    \caption{Dynamic response of the gFHR model for 5\% to 40\% power drops.
    Subplots A to E: controlled response of the reactor power along with the reactivity components (external reactivity insertion, moderator \& fuel feedback, coolant feedback, and total reactivity).
    Subplots F to K: controlled response of core inlet and outlet temperatures through primary and secondary circuit pump actuation.
    Subplots L and M: uncontrolled response of secondary-side heat exchanger inlet and outlet temperatures.
    Outputs of PID controllers are annotated with a ``\textcolor{textred}{PID($\cdot$) Output}''.
    }
    \label{fig:gfhr_sch}
\end{figure}

In \cref{fig:gfhr_sch}, the plant response is presented for multiple transients, ranging from a 5\% to 40\% drop in nominal power.
As the reactor power drops (\cref{fig:gfhr_sch}A), there is a large reactivity insertion due to the moderator density \& fuel Doppler feedback (\cref{fig:gfhr_sch}C) and a lesser coolant density feedback (\cref{fig:gfhr_sch}D).
At the peak of the 40\% power drop, almost -0.18 \$ of external reactivity insertion (\cref{fig:gfhr_sch}B) is required. 
The peak total reactivity ($\rhosig=\rhoext+\rhom+\rhoc$, \cref{fig:gfhr_sch}E) is approximately -1.65 \textcentoldstyle. 
As reactor power drops, two PID controllers decrease the head supplied by the primary and secondary pumps (\cref{fig:gfhr_sch}G and J, respectively), decreasing the primary and secondary mass flow rates (\cref{fig:gfhr_sch}H and K, respectively).
Despite the initial undershoots, the response of core outlet and core inlet temperature (\cref{fig:gfhr_sch}F and I, respectively) remains within \SI{1}{\celsius}. 
However, there are no active controllers to keep the secondary-side temperatures constant during the transient.
The inlet and outlet temperatures at the secondary-side heat exchanger (\cref{fig:gfhr_sch}L and M, respectively) reach a new equilibrium value.
For a 40\% power drop, $\Tsi$ decreases by approximately \SI{2.2}{\celsius}, and $\Tso$ increases by approximately \SI{6.0}{\celsius}.
Since the gFHR system is underactuated, \ie the available actuators are fewer than the process variables that we would like to control, a \textit{supervisory layer is necessary} to constrain the variations of $\Tsi$ or $\Tso$.

\section{Methods}\label{sec:methods}

To address the objectives of this work, several algorithms were integrated.
First, the algorithm chosen to enforce constraints, the Reference Governor, is described in \cref{subsec:rg}. 
The algorithm to identify the state-space representation matrices, Dynamic Mode Decomposition with Control, is described in \cref{subsec:si}.
The feature selection algorithm utilized to define the constituents of the dynamics model, Sequential Feature Selection, is described in \cref{subsec:statesel}.
The feature selection procedure indicated that some unobserved variables (\eg delayed neutron precursor concentrations) were optimal candidates for the dynamics model.
Thus, a non-linear Kalman Filter, the Unscented Kalman Filter, was implemented, discussed further in \cref{subsec:kf}.
Finally, the composition of the supervisory control system architecture is outlined in \cref{subsec:arch}.

\subsection{Constraint Enforcement}\label{subsec:rg}
\begin{figure}
    \centering
    \captionsetup{width=.65\linewidth}
    \includegraphics[width=0.65\linewidth]{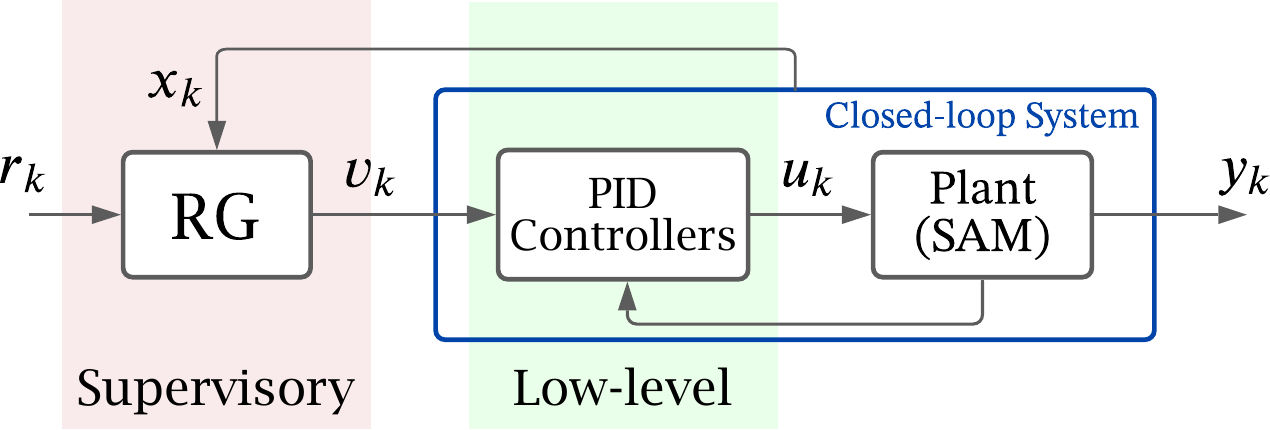}
    \caption{Block diagram visualizing the topology of the proposed supervisory control system.
    The placement of the supervisory and low-level layers is highlighted.
    Direct access to the underlying closed-loop system is always available by bypassing the supervisory layer. 
    }
    \label{fig:RG}
\end{figure}

The Reference Governor (RG) \cite{Bemporad1994} is the constraint enforcement algorithm utilized in this work.
The RG is an \textit{add-on} control scheme that enforces \textit{point-wise in time} constraints.
The terminology \textit{add-on} reflects the nature of its intended use: the algorithm is applied to a system that already has an existing closed-loop feedback system (\eg PID controllers for regulation).
The add-on feature is important as an operator may choose to manually override the RG, while maintaining controllability to maneuver the plant manually.
The terminology \textit{point-wise in time} reflects the capability to update constraint bounds at each discrete time-step.
This feature allows addressing process drift scenarios, \eg degradation phenomena.
The application of the RG is visualized in \cref{fig:RG}.
The RG algorithm is described in detail in the rest of the section.
A full review of the RG algorithm family, including a comparison against Model Predictive Control (MPC) is provided in \citep[\S6.1]{Garone2017}).
Unlike the MPC, the RG is formulated with constraint enforcement as the primary objective to fulfil during supervisory control.
The RG acts as an agent that checks whether the requested setpoint variation violates pre-defined constraint(s) on the system.
The RG accomplishes this quantitatively by utilizing a surrogate model of the system, and propagating it forward in time.

To gain some intuition, the basic form of the RG algorithm, the Scalar RG (SRG) is explained.
At each time-step $k\in\mathbb{Z}_+$, the SRG receives a reference input, $r_k\in\mathbb{R}^m$, where $m$ is the number of inputs.
Depending on the current state of the system, expected evolution, and user-defined constraints, an admissible input $v_k\in\mathbb{R}^m$,
\begin{equation}
    v_k = v_{k-1} + \kappa_k \left(r_k - v_{k-1}\right)~,
    \label{eq:rg1}
\end{equation}
is sent to the low-level controllers.
In \cref{eq:rg1}, $\kappa_k\in[0,1]$ is a \textit{scalar} that governs admissible changes to the inputs, such that a complete rejection, $v_k=v_{k-1}$, acceptance, $v_k=r_k$, or an intermediate change, $v_{k-1}\leq v_k\leq r_k$, is possible.
After determining $\kappa_k$, the output of the RG, $v_k$ is sent to the existing controllers.
For example, a PID controller would receive a component of $v_k$ as a reference set-point for its MV, and according to its paired MV's current value, issue a control action $u_k$ to the CV.

Next, the steps to evaluate $\kappa_k$ are summarized.
The RG relies on a state-space representation,
\begin{align}
    x_{k+1} & = Ax_k + Bu_k~, \label{eq:state}\\
    y_k & = Cx_k + Du_k~, \label{eq:outp}
\end{align}
where $x_k\in\mathbb{R}^n$, $u_k\in\mathbb{R}^m$, and $y_k\in\mathbb{R}^p$ are vectors at step $k$ representing states, inputs, and outputs, respectively.
The matrix $A\in\mathbb{R}^{n\times n}$ is the state matrix.
The matrix $B\in\mathbb{R}^{n\times m}$ is the input matrix.
The matrix $C\in\mathbb{R}^{p\times n}$ is the output matrix.
The matrix $D\in \mathbb{R}^{p\times m}$ is the control pass-through matrix.
Using the system model defined in \cref{eq:state,eq:outp}, constraints are imposed on the output variables, $y_k\in Y$, where $Y$ is defined by a set of linear inequalities.
Constraints may be imposed on $x_k$, or $u_k$.
To ensure constraints are not violated, we define the maximal output admissible set (MOAS), or admissible region.
The admissible region is the set of all $x_k$, and constant input $\Tilde{v}$, such that,
\begin{equation}
    O_\infty = {(x_k, \Tilde{v}) : y_{t+k}\in Y,~v_{t+k}=\Tilde{v},~\forall k \in \mathbb{Z}^+}~.
    \label{eq:moas}
\end{equation}
The $k\rightarrow \infty$ assumption of $O_\infty$ is generally relaxed to a suitably large finite horizon, $T$.
Thus, at each time-step $k$, the SRG algorithm determines the admissible $\Tilde{v}$, and therefore $\kappa_k$, such that the system remains in $O_\infty$.
The $O_\infty$ is constructed by evaluating \cref{eq:state,eq:outp} for each $k=0,1...T$ time steps.
In \cref{fig:moas-abs}, the optimal operating region was qualitatively introduced to constrain the plant responses during operational transients.
Now, the optimal operating region can be quantitatively represented by the admissible region.

The RG formulation can also account for process and output disturbances.
To this aim, \cref{eq:state,eq:outp} are modified,
\begin{align}
    x_{k+1} & = Ax_k + Bu_k +B_w w_k~, \label{eq:state_w}\\
    y_k & = Cx_k + Du_k + D_w w_k~, \label{eq:outp_w}
\end{align}
where $B_w$ and $D_w$ are process and output disturbance matrices.
Additionally, $w_k$, is a scalar representing a bounded disturbance, $w_k\in W$, where W is a compact set containing the origin, \eg $W=[-1, 1]$.
The definition of the admissible region in \cref{eq:moas} is modified such that,
\begin{equation}
    O_\infty = {(x_k, \Tilde{v}) : \tilde{y}_{t+k}\subseteq Y,~v_{t+k}=\Tilde{v},~ \forall k \in \mathbb{Z}^+}~,
    \label{eq:moas_w}
\end{equation}
where $\tilde{y}_k$ is a set-valued prediction of the output accounting for all possible disturbances.
Because \cref{eq:state_w,eq:outp_w} are linear, the set of inequalities imposed ($Y$) are modified to account for these disturbances \citep[\S2.2]{Garone2017}.

A drawback of the SRG formulation is that $\kappa_k$ is a scalar.
Therefore, if $v_k-r_k$ is multi-dimensional and $\kappa_k<1$, input variation in all dimensions is bounded.
In \citep{dave2022numerical}, constraint enforcement for multiple inputs was demonstrated for a molten salt loop, using the Command Governor, a variant of the RG which selects $v_k$ by solving a quadratic program.
In this work, as the high-level input is a scalar, the SRG is sufficient.

\subsection{System Identification}\label{subsec:si}

To predict system response and detect potential violations of imposed constraints, the RG algorithm needs a state-space representation of the system.
In this work, Dynamic Mode Decomposition with Control (DMDc) \cite{Proctor2016} is the system identification algorithm that is used to define the $A$ and $B$ matrices in \cref{eq:state}.
DMDc is an extension of Dynamic Mode Decomposition (DMD), a method that originated in the fluids community to extract spatial-temporal coherent structures from experimental data.
Qualitatively, the DMDc methodology combines Singular Value Decomposition (dimensionality reduction) with a Fourier transform (finding dominant patterns in time).
DMDc is a data-driven method relying on a temporal sequence of data, of length $L$.
The history of states is grouped into matrices,
\begin{align}
    X'&= \begin{bmatrix}
    \vline & \vline & & \vline \\
    x_1 & x_2 & \cdots & x_L\\
    \vline & \vline & & \vline
    \end{bmatrix} ,\\
    X&= \begin{bmatrix}
    \vline & \vline & & \vline \\
    x_0 & x_1 & \cdots & x_{L-1}\\
    \vline & \vline & & \vline
    \end{bmatrix} ,
    \label{eq:snaps}
\end{align}
and the history of inputs into matrix,
\begin{equation}
    \Upsilon = \begin{bmatrix}
    \vline & \vline & & \vline \\
    u_0 & u_1 & \cdots & u_{L-1} \\
    \vline & \vline & & \vline
    \end{bmatrix} .
    \label{eq:snapu}
\end{equation}
Each transient was \SI{2000}{\second} long with a time-step of \SI{0.2}{\second}. The evolution of the system is represented by,
\begin{align}
    X' &= AX + B\Upsilon~,~\mathrm{or}, \\
    X' &= G\Omega,
\end{align}
where $G=\left[A~B\right]$ and $\Omega=[X^\intercal \Upsilon^\intercal]^\intercal$.
To find the reduced order approximations of matrices $A$ and $B$ \cite[\S3.3]{Proctor2016}, solve for $G$,
\begin{equation}
    G = X'\Omega^\dagger~,
\end{equation}
where the superscript $\dagger$ represents the psuedoinverse of a matrix.

\subsection{State Selection}\label{subsec:statesel}

The RG algorithm requires that \cref{eq:state} accurately represents the dynamics of the system. 
To this aim, optimally selecting the state variables, $\bx_k$, is critical.
However, given the large number of candidate process variables (102), a systematic approach is necessary.
In this work, the Sequential Forward Floating Selection (SFFS) feature selection algorithm \citep{pudil1994floating} is used.
The psuedocode for the SFFS algorithm is listed in \cref{alg:sfs}.
First, the set of state variables, $H$, that must always be included are defined.
Next, the set of all possible supplementary states, $Y$, is defined.
There were a total of 8 states defined in $H$, therefore there were a total of 94 supplemental states to choose from. 
The candidate process variables were mass flow rates, temperature and pressure measurements, delayed neutron precursor concentration (discussed further in the next section), heat transfer rates across heat exchanger and steam generator ($\qhx$ and $\qsg$), pump heads, and PID controller components.
After defining $H$ and $Y$, the set of supplemental candidates, $X_k$, is incrementally expanded until the maximum number of variables, $k_T$, is reached.
The optimal candidate at each iteration, $k$, is determined by assessing the objective function,
\begin{equation}
    J(X_k) = \frac{1}{N}\sum_n^N \left(\frac{1}{2} r_\mathrm{tr}(x_i) + \frac{1}{2} r_\mathrm{te}(x_i)\right) ~,
    \label{eq:sfsobj}
\end{equation}
where $x_i \in H + X_k$, $r$ is a residual function that is assessed on training ($r_\mathrm{tr}$) and test ($r_\mathrm{te}$) data, and $N$ is the total number of unique trajectories used for evaluating $r$.
The mean square error and coefficient of determination ($R^2$) were used as candidate residual functions to quantify the accuracy of DMDc \vs SAM data.
Due to the significant variation in scales (\eg pressure \vs temperature), the mean square residual components were individually normalized.
However, the coefficient of determination metric still provided superior results.

\begin{algorithm}
\caption{Sequential Forward Floating Selection}\label{alg:sfs}
\begin{algorithmic}
\Inputs{
$H=\{h_1, h_2, ..., h_d\}$ \\
$Y=\{y_1, y_2, ..., y_e\}$ \\
$k_T$ }
\Initialize{
$k \gets 1$ \\
$X_0 \gets \emptyset$
}
\While {$k \neq k_T$}
\State $x^+ \gets \argmin_x J(H + X_k + x)$, where $x\in Y - X_k$
\State $X_{k+1} \gets X_k + x^+$
\State $k \gets k + 1$
\If{floating}
\State $x^- \gets \argmin_x J(H + X_k - x)$, where $x\in X_k$
\If{$J(X_k -x) > J(X_k)$}
\State $X_{k-1} \gets X_k - x^-$
\State $k \gets k -1$
\EndIf
\EndIf
\EndWhile
\end{algorithmic}
\end{algorithm}

\begin{figure}
    \centering
    \captionsetup{width=.8\linewidth}
    \includegraphics[width=0.8\linewidth]{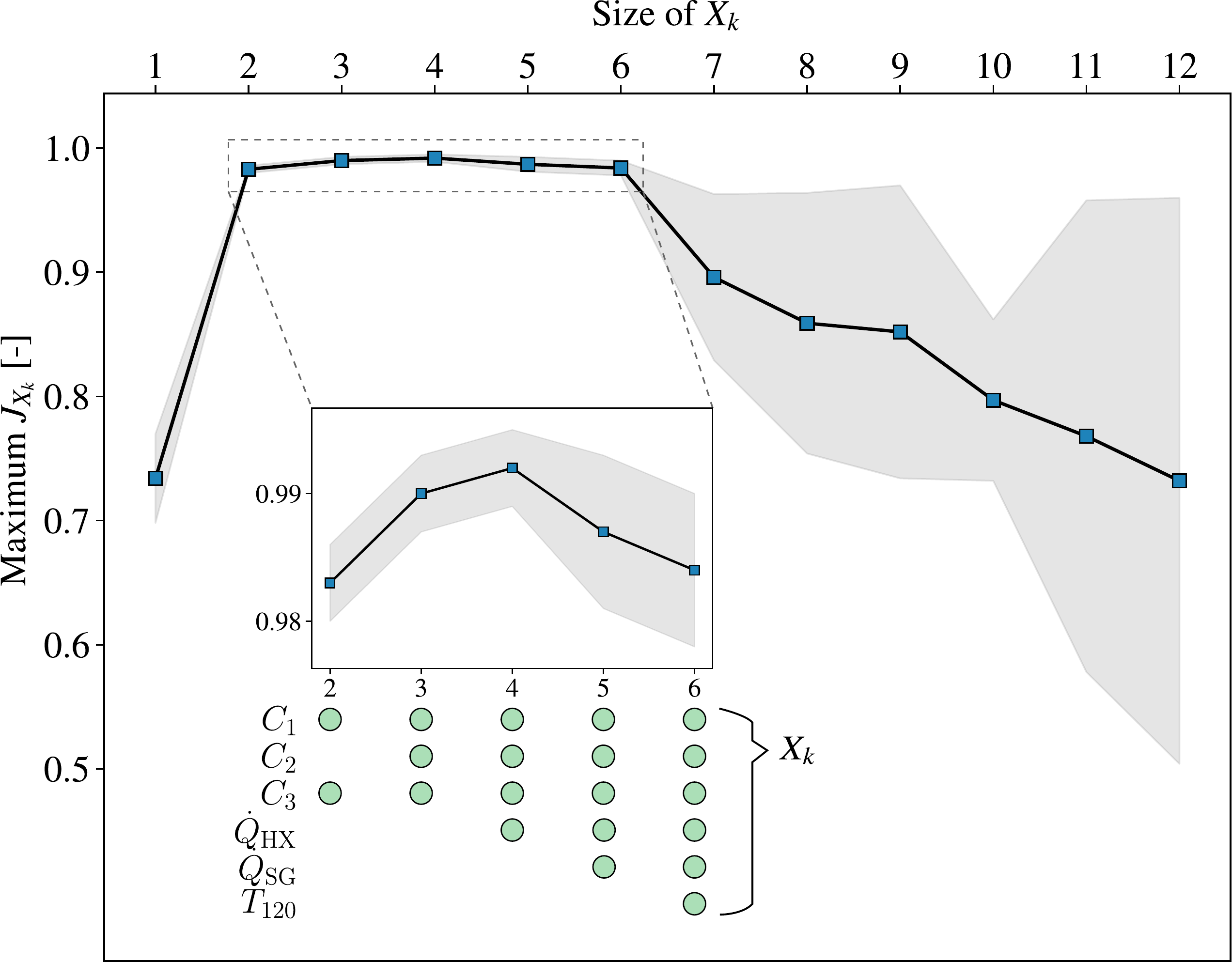}
    \caption{To determine the optimal composition of $X_k$, the SFFS algorithm (\cref{alg:sfs}) is used.
            For each size of $X_k$, the maximum achievable objective value, $J_{X_k}$ (\cref{eq:sfsobj}), is presented.
            The shaded regions indicate one standard deviation across 3 cross-validation sets.
            Inset is the performance for sets $X_k: k\in\{2,3,...,6\}$.
            Below the inset, the optimal composition for each $X_k$ is presented.
        }
    \label{fig:sfs}
\end{figure}

The results from applying the the SFFS algorithm are discussed next.
The 8 state variables in $H$ were $\mfp$, $\mfs$, $\Tci$, $\Tco$, $\Pci$, $\Pco$, $\Tsi$, and $\Tso$ (identified in \cref{fig:gfhr-pid}).
A total of 22 unique trajectories were generated for training.
At each SFFS iteration, the DMDc algorithm is applied to the training data and then the objective function is calculated, \cref{eq:sfsobj}.
The maximum objective function value at each SFFS iteration and its spread across cross-validation sets is presented in \cref{fig:sfs}.
The objective function drastically increases after introducing two delayed neutron precursor groups ($C_1$ and $C_3$), and climaxes once the heat transfer rate across the heat exchanger ($\qhx$)  is added.
Increasing $X_k$ beyond $k=6$ results in deteriorating performance, with very high variance across cross-validation datasets.
These results indicate that there is a quantitatively optimal point achieved at $k=4$.
However, at $k=5$, addition of the heat transfer rate across the heat exchanger ($\qsg$) led to more conservative predictions of temperatures, with similar regression performance.
Therefore, $C_1$, $C_2$, $C_3$, $\qhx$, and $\qsg$ were chosen as the complimentary state variables to $H$.

\subsection{State Observer} \label{subsec:kf}

The variables that were added to $H$ included unobserved variables, such as the delayed neutron precursor (DNP) group concentration.
These variables cannot be directly measured.
To estimate their values, state observers such as the Kalman Filter (KF) must be used.
Recent work by \citet{Bhatt2013} studied the experimental observation of reactivity using the Extended Kalman Filter (EKF).
While the KF is restricted to linear dynamical systems, the EKF observes non-linear dynamical systems.
The major disadvantage of the EKF is that it requires explicit definition of Jacobians.
Instead, in this work, the Unscented Kalman Filter (UKF) is adopted \cite{julier1997new}.
The UKF samples the non-linear process model, and approximates the Jacobians.
Therefore, to use the UKF, a non-linear process model representing the nuclear reactor dynamics is required.

The non-linear dynamical model used by SAM to simulate nuclear reactor dynamics is presented next.
The 6-group point kinetics equation (PKE) is defined by,
\begin{align}
    \frac{dn}{dt} &= \frac{\rho-\beta}{\Lambda}n + \sum_{i=1}^{6} \lambda_i C_i~, \label{eq:pwr} \\
    \frac{dC_i}{dt} &= \frac{\beta_i}{\Lambda}n - \lambda_i C_i~, \label{eq:dnp}
\end{align}
where $n$ is the neutron population (proportional to the reactor power), $\rho$ is the reactivity, $\beta$ is the total DNP fraction ($\beta = \sum_{i=1}^{6}\beta_i$), $\Lambda$ is the prompt neutron generation time, $\lambda_i$ is the $i$-th DNP group decay constant, and $C_i$ is the $i$-th delayer neutron precursor group concentration.
In \cref{eq:pwr,eq:dnp}, $\beta_i$, $\lambda_i$, and $\Lambda$ are constants determined a priori via high-fidelity neutronics models (\eg Monte Carlo neutron transport) and depend on the reactor design.
To model reactivity, a first order power series \citep{venerus1970} is chosen,
\begin{equation}
    \rho = \alpha + \omega t~, \label{eq:rhomod}
\end{equation}
where $\alpha$ represents an instantaneous increase in $\rho$, and $\omega$ represents the slope of reactivity increase.
\cref{eq:pwr,eq:dnp,eq:rhomod} can be written in the state space form,
\begin{equation}
    \frac{d\bx}{dt} = \bA \bx + \bB~, \label{eq:contss}
\end{equation}
where,
\begin{align}
\bA &=
\begin{bmatrix}
-\beta/\Lambda   & \lambda_1 &  \lambda_2 & \lambda_3 & \lambda_4 & \lambda_5 & \lambda_6 & 0 & 0 \\
-\beta_1/\Lambda & -\lambda_1&  0         &         0 &         0 &         0 &         0 & 0 & 0 \\
-\beta_2/\Lambda &         0 &  -\lambda_2&         0 &         0 &         0 &         0 & 0 & 0 \\
-\beta_3/\Lambda &         0 &  0         & -\lambda_3&         0 &         0 &         0 & 0 & 0 \\
-\beta_4/\Lambda &         0 &  0         &         0 & -\lambda_4&         0 &         0 & 0 & 0 \\
-\beta_5/\Lambda &         0 &  0         &         0 &         0 & -\lambda_5&         0 & 0 & 0 \\
-\beta_6/\Lambda &         0 &  0         &         0 &         0 &         0 & -\lambda_6& 0 & 0 \\
               0 &         0 &          0 &         0 &         0 &         0 &         0 & 0 & 1 \\
               0 &         0 &          0 &         0 &         0 &         0 &         0 & 0 & 0
\end{bmatrix}~,~\mathrm{and}, \label{eq:A} \\
\bB &=
\begin{bmatrix}
\frac{\rho n}{\Lambda} & 0 & 0 &0 &0 &0 &0 &0 &0
\end{bmatrix}^T~. \label{eq:B}
\end{align}
To proceed with the implementation, \cref{eq:contss} must be transformed from the continuous to the discrete space.
We use the integral approximation method to discretize the continuous equations, and assume a zero-order hold for the constants in between the sampling period.
The resulting equations are,
\begin{equation}
    \bx_{k+1} = e^{\bA\Delta t}\bx_k + \int_0^{\Delta t}e^{\bA t}\bB dt~,
    \label{eq:xdisc}
\end{equation}
where $\bx_k$ is the state at time-step $k$, and $\Delta t$ is the size of the discrete time-step.
To use the UKF, \cref{eq:A,eq:B,eq:xdisc} are provided as the discrete state transition function.
The measurement function returns $n$ to the filter (in an operating reactor, this measurement would come from ex-vessel neutron detectors).
Van der Merwe's algorithm \citep{van2004sigma} was used to generate Sigma Points (with parameters $\alpha=0.001,~\beta=2.0,~\kappa=0$).

\begin{figure}
    \centering
    \includegraphics[width=0.925\linewidth]{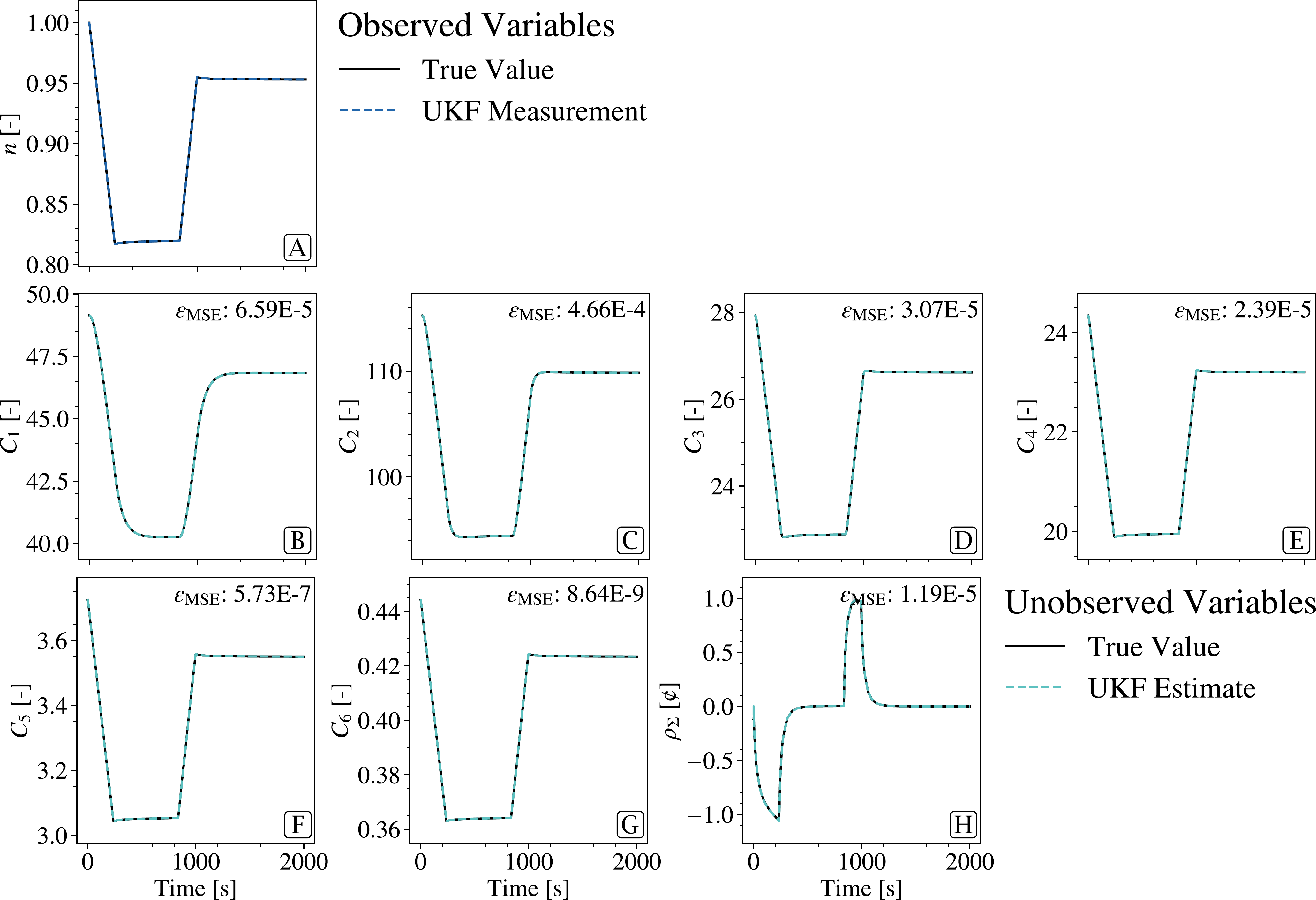}
    \caption{Application of the UKF algorithm for estimating unobserved variables in the non-linear 6-group PKE.}
    \label{fig:ukf-no-noise}
\end{figure}

\begin{figure}
    \centering
    \includegraphics[width=0.925\linewidth]{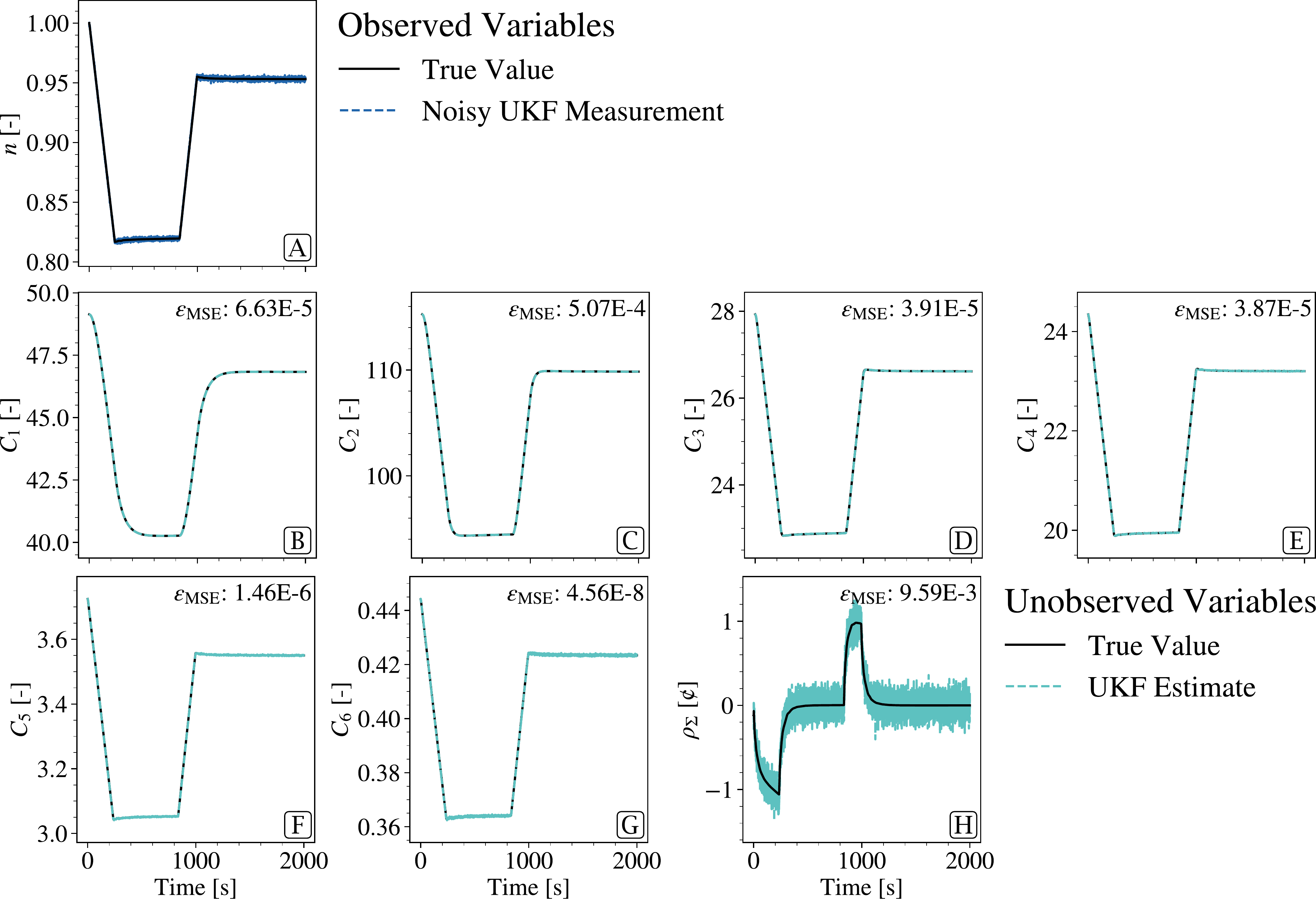}
    \caption{Application of the UKF algorithm with white noise added to the power measurement.}
    \label{fig:ukf-noisy}
\end{figure}

The application of the UKF algorithm during operational transients is discussed next.
In \cref{fig:ukf-no-noise}, a transient featuring a reduction to 82.5\% of nominal power, following an increase back to 95\% power is presented.
The agreement between the UKF estimates and the true values for all DNP groups (\cref{fig:ukf-no-noise}B to G) and the total reactivity insertion (\cref{fig:ukf-no-noise}H) is satisfactory, as quantified by the mean-square error.
To assess the performance in noisy environments, white gaussian noise with $\sigma=0.001$ was added to the measurements of $n$.
In \cref{fig:ukf-noisy}, the UKF estimates during the same transient, except with noise, are presented.
While the error in predicting the DNP remains consistently low (\cref{fig:ukf-noisy}B to G), there is a significantly larger error in predicting the total reactivity insertion (\cref{fig:ukf-noisy}H).
This result indicates that while DNP group concentration would be a good candidate for state variables, in noisy environments, the total reactivity feedback would be a poor choice or would require further denoising.

\subsection{Proposed Control System Architecture}\label{subsec:arch}

\begin{figure}
    \centering
    \captionsetup{width=.70\linewidth}
    \includegraphics[width=0.65\linewidth]{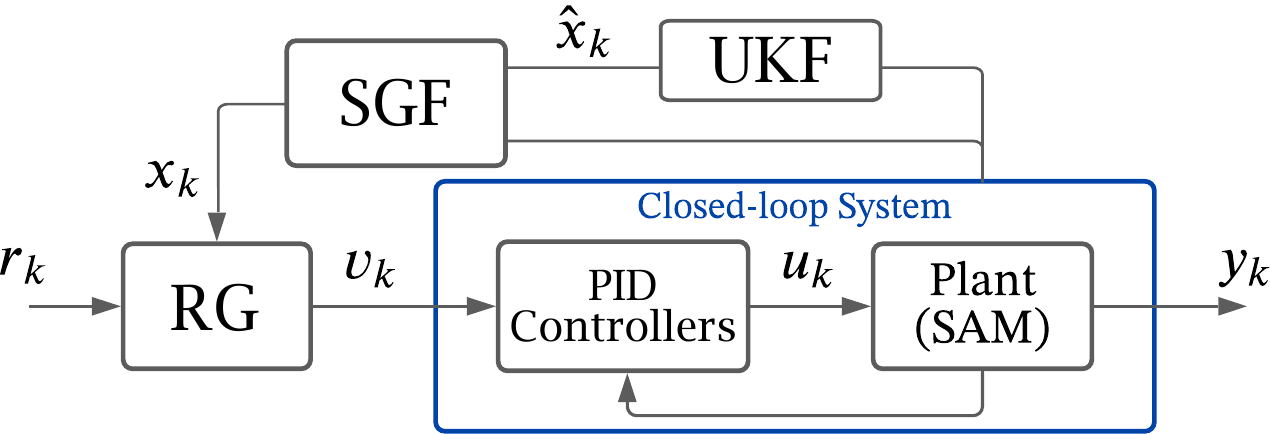}
    \caption{Block diagram visualizing the topology of the Reference Governor (RG) in conjunction with the Unscented Kalman Filter (UKF).
             A partial subset of the full state set, $x_k$, is predicted by the UKF algorithm, $\hat{x}_k$.
             Under the presence of noise, all outputs are denoised using the Savitzky-Golay Filter (SGF).
    }
    \label{fig:RG_ukf}
\end{figure}

The selection of state variables indicated that the inclusion of DNPs is important for an accurate DMDc model of the gFHR.
To infer DNPs evolution, the non-linear state observer must be embedded.
To incorporate the UKF into the control system, the configuration proposed in \cref{fig:RG} needs to be modified to that in \cref{fig:RG_ukf}.
The UKF will require a single output from the plant, \ie the power measurement.
A subset of the UKF predictions chosen by the feature selection algorithm, $C_{1\rightarrow 3}$, will update their respective value in the overall state of the plant.
The notation $x_k$ represents the entire set of state variables, whereas $\hat{x}_k$ represents a subset predicted by the UKF.
An additional block, ``SGF'', is upstream of the RG block.
The Savitzky-Golay Filter (SGF) is a denoising filter that will be introduced further in \cref{subsec:res_noise}.

\FloatBarrier

\cleardoublepage
\section{Supervisory Control of an Advanced Reactor}\label{sec:results}

This section will focus on applying the supervisory control system to the 100 \% to 60 \% load-follow transient.

\subsection{Accuracy of DMDc Models}

\begin{figure}
    \centering
    \includegraphics[width=\linewidth]{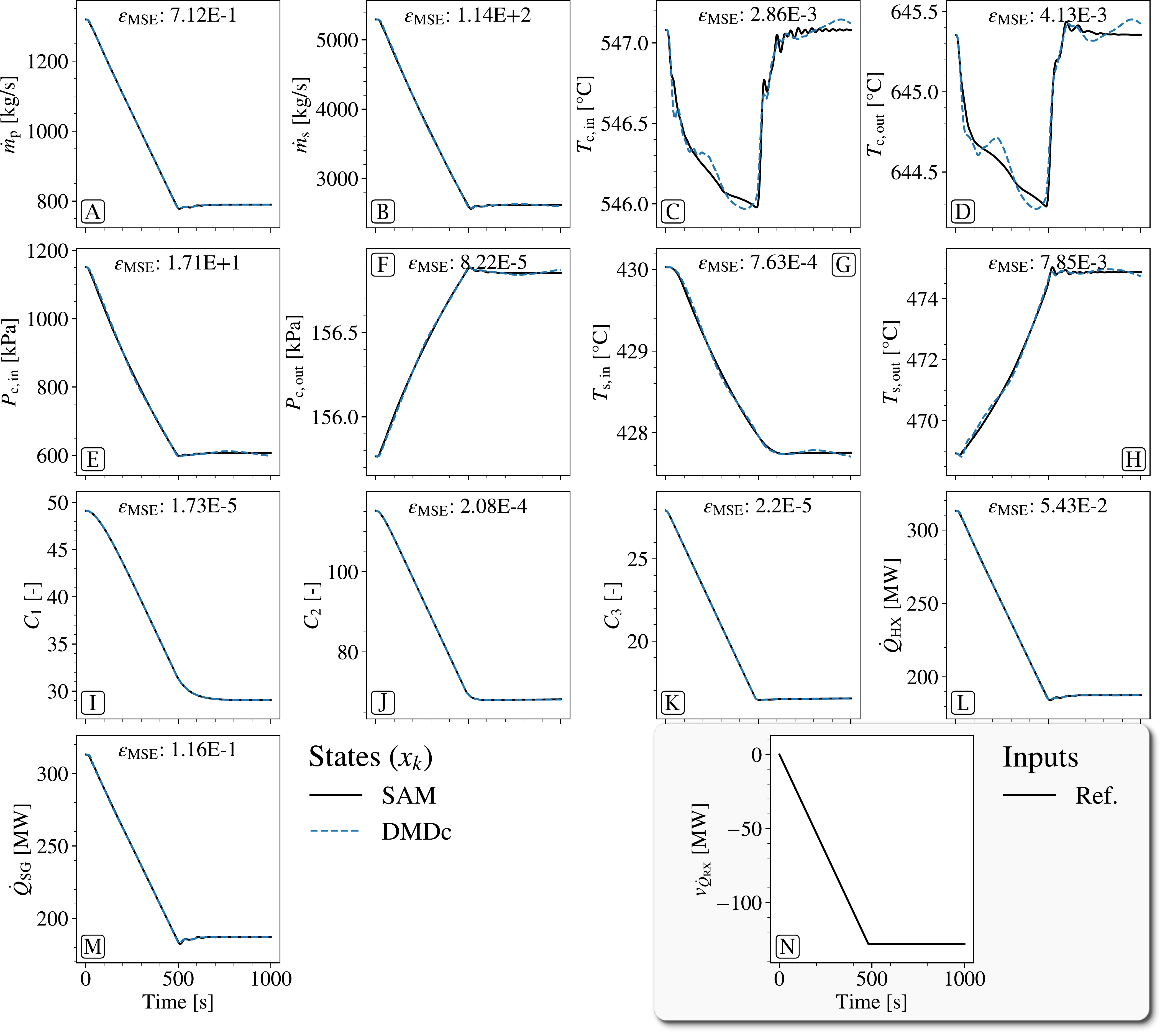}
    \caption{
        Assessment of the DMDc model accuracy during the 40 \% power drop.
        Gray box presents the reference input that is shared between all results in this section.
        Remaining plots display contrast of the propagation of states between the SAM model and the DMDc model.
        The mean square error, $\varepsilon_\mathrm{MSE}$, is inset in the subplot of each state.}
    \label{fig:res-sysid}
\end{figure}

First, the accuracy of the forward dynamics model, the DMDc model, is discussed.
In \cref{fig:res-sysid}, the responses of the state variables predicted by the DMDc model are presented. 
Quantitatively, there is a very low mean square error in prediction of almost all variables.
The only variables where the quantitative and qualitative propagation could potentially improve are the core inlet and outlet temperatures (\cref{fig:res-sysid}C and D, respectively).
There is a physical reason that can explain why these variables are more difficult to model by adopting a linear system, \ie \cref{eq:state}.
As noted in \cref{sec:gfhr}, the gFHR model includes RCCS.
The RCCS system is passively driven by buoyancy, which is directly impacted by the heat loss through the reactor vessel.
The relationships between power generated, reactor vessel heat loss, and RCCS are not linear, and so is the heat transfer rates out of the reactor vessel during load-follow operation.
The non-linearity would impact temperature propagation in the primary side, \ie $\Tci$ and $\Tco$.
The secondary side temperatures (\cref{fig:res-sysid}G and H) are also impacted, but to a much lesser extent.

For the present purposes, the DMDc model provides a good approximation of the plant dynamics.
Thus, the RG algorithm can use the model identified using the DMDc algorithm to enforce constraints.
It is important to note that while the RG algorithm uses the DMDc model to approximate $O_\infty$, the resulting control signal ($v_k$ in \cref{fig:RG_ukf}) is sent to the SAM model.
In other words, while a linear surrogate is used to approximate the bounds of the admissible region in \cref{fig:moas-abs}, the final control action is sent to a non-linear, high-fidelity model of the FHR. 
Therefore, the proceeding results are all outputs from SAM -- not the DMDc model.

\subsection{Application of the Supervisory Layer}\label{subsec:rg_res}

The application of the supervisory layer to enforce constraints is presented next.
The selected state variables (\cref{subsec:statesel}), the reference input to the RG, and the constrained variables are tabulated in \cref{tab:rg_vars}.
\begin{table}%
    \centering
    \caption{%
    State, input and constrained variables assigned to the RG algorithm.
    }
    \begin{tabular}{ccc}%
        \begin{tabular}[t]{l}
            \hline\T\B
            States -- $x_k$\\
            \hline\T\B
            $\mfp$, $\mfs$, \\
            $\Tci$, $\Tco$, \\
            $\Pci$, $\Pco$, \\
            $\Tsi$, $\Tso$, \\
            $C_1$, $C_2$, $C_3$, \\
            $\qhx$, $\qsg$
        \end{tabular} &
        \begin{tabular}[t]{l}
            \hline\T\B
            Inputs -- $r_k$\\
            \hline\T\B
            $\qrxref$
        \end{tabular} &
        \begin{tabular}[t]{l}
            \hline\T\B
            Constrained Variables\\
            \hline\T\B
            $\Tso \leq \Tsomax$\\
            $\Tsi \geq \Tsimin$\\
            $\delta\vqrx \leq \delta\vqrxmax$
        \end{tabular}
    \end{tabular}
    \label{tab:rg_vars}
\end{table}
The state values, $x_k$, are used as an initial condition at each time step $k$ by the RG to determine if a constraint would be violated due to a change in the reference input $r_k$.
Note that the reference input, $r_k$, only consists of the reactor power ($\qrxref$), as both pumps are used for regulation only.
In the control strategy (\cref{subsec:strategy}), it was decided that $\Tsi$ and $\Tso$ are appropriate candidates to constrain during load-follow.
Thus, the constraints applied are,
\begin{equation*}
  \Tsimin=\begin{cases}
       \SI{428.23}{\celsius}, &t\leq \SI{700}{\second}\\
       \SI{427.98}{\celsius}, &t> \SI{700}{\second}
    \end{cases}
    \label{eq:constr}
\end{equation*}
and $\Tsomax = \SI{474.43}{\celsius}$. The value for $\Tsimin$ is time-dependent to assess the adaptability of the supervisory layer.
Additionally, a constraint is applied to the rate of variation of reactor power between timesteps, $\delta\vqrx$.
The maximum value is set to $\delta\vqrxmax = \SI{16}{\mega\watt\per\minute}$ (or 5\% nominal power per minute).
This constraint is imposed to prevent large reactivity insertions.

\begin{figure}
    \centering
    \includegraphics[width=\linewidth]{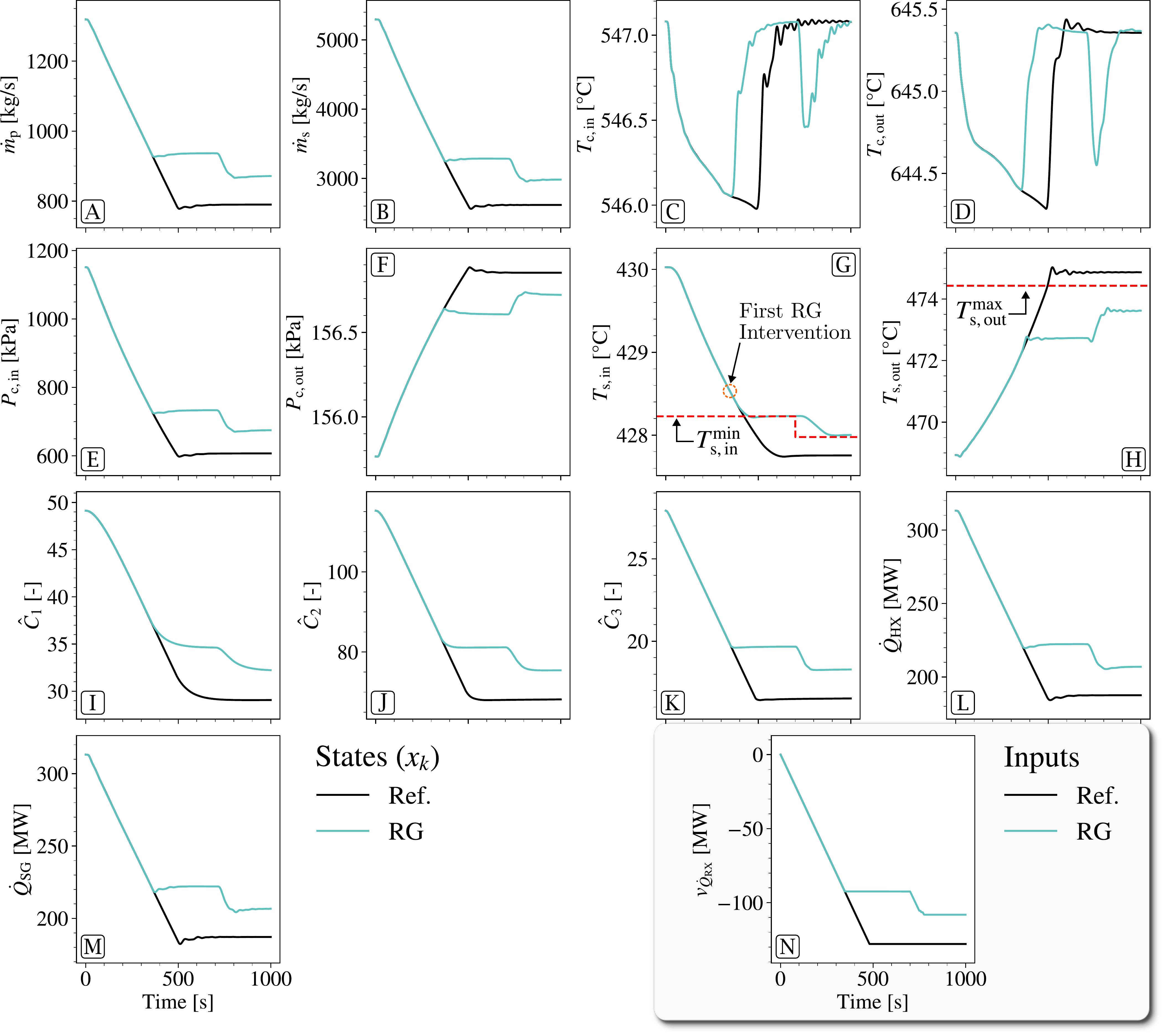}
    \caption{Application of the supervisory layer to enforce constraints on the heat exchanger secondary-side inlet and outlet temperatures.
             Gray box contrasts the reference requested input value (``Ref.'') against the admitted input value (``RG'').
             Remaining plots display changes in state propagation as the constraints on $\Tsi$ and $\Tso$ are enforced by the RG.
             Plots of $\Tsi$ and $\Tso$ display the constraints provided to the RG algorithm.
            }
    \label{fig:res-rg}
\end{figure}

The results of applying the supervisory layer during the load-follow transient are presented in \cref{fig:res-rg}.
Throughout the transient, the constraint applied to $\Tso$ (\cref{fig:res-rg}H) is not active.
Thus, the following discussions will focus on $\Tsi$ (\cref{fig:res-rg}G).
In the Inputs subplot (\cref{fig:res-rg}N), it is shown that the admitted change in setpoint sent to the low-level controllers varies from the reference trajectory.
An intervention preventing the reactor power from further decreasing occurs at approximately \SI{347}{\second}.
In the $\Tsi$ subplot, the point at which this occurs is annotated as ``First RG Intervention'', and it occurs before the constraint threshold is violated.
However, as time progresses, the state of the system continues to evolve and eventually $\Tsi\approx\Tsimin$.
Therefore, there is a lag between the first intervention in change of reactor power occurring and the value of the constrained process variable settling around the constraint value.
Physically, there is a transport lag between changes in the reactor power and changes in temperature measurements in the secondary-side.
This delay is accurately captured by the DMDc model and aids in enforcing the constraint via the RG.

In \cref{subsec:rg}, the point-wise in time capability of the RG was outlined.
During the transient, the $\Tsimin$ constraint is updated at \SI{700}{\second} to be \SI{0.25}{\celsius} lower.
The RG algorithm immediately takes advantage of this additional region available to decrease the reactor power further to meet the reference setpoint, while adhering to the $\delta\vqrx \leq \delta\vqrxmax$ constraint.
A second RG intervention occurs afterwards.
Again, there is a delay in the intervention occurring and the $\Tsi$ state value settling around $\Tsimin$.
The capability of updating constraints during transients is important when the control system is coupled to other modules (\eg updating constraints due to a diagnosis algorithm).
The resulting supervisory control system is adaptable, as it allows addressing constraint updates from auxiliary algorithms.

\FloatBarrier
\subsection{Accommodating Sensor Noise}\label{subsec:res_noise}

The RG formulation accommodates process and output disturbances.
The former addresses modeling uncertainties \cref{eq:state}; the latter addresses measurements uncertainties, \cref{eq:outp}.
Both equations are then modified, \cref{eq:state_w,eq:outp_w}), to account for disturbances.
The RG addressing these disturbances is known as ``Robust RG''.
In this section, the capability to account for output noise is exclusively explored.
The capability is incorporated due to a pragmatic need -- ultimately, the proposed control system would be applied to physical systems that are monitored with sensors subject to aleatoric noise.
To simulate output disturbances, Gaussian white noise is added to all outputs, tabulated in \cref{tab:outpw}.
\begin{table}
    \centering
    \caption{Gaussian white noise added to output measurements during the load-follow transient.}
    \label{tab:outpw}
    \begin{tabular}{c c}
        \hline\T\B
        Output & $3\sigma$ disturbance \\
        \hline\T\B
        $\mfp$, $\mfs$ & \SI{15}{\kilo\gram\per\second} \\
        $\Tci$, $\Tco$, $\Tsi$, $\Tso$ & \SI{0.5}{\celsius} \\
        $\Pci$, $\Pco$ & \SI{0.1}{\kilo\pascal} \\
        $\qhx$, $\qsg$ & \SI{1.0}{\kilo\watt} \\
        $n$ (\cref{eq:pwr}) & 0.003
    \end{tabular}
    \vskip -1.5em
\end{table}
The output disturbances are assigned to the matrix $D_w$ in \cref{eq:outp_w}.
Noise is also added to $n$ sent to the UKF, as in \cref{fig:ukf-noisy}, indirectly adding noise to $\hat{C}_{1\rightarrow 3}$.
All noisy measurements are processed by the Savitzky-Golay filter (SGF) \citep{savitzky1964smoothing}, with a 299 step window length, a third order polynomial fit, and derivatives disabled. 
The final configuration of the supervisory control system was presented in \cref{fig:RG_ukf}.
The UKF provides a prediction for $\hat{C}_{1\rightarrow 3}$ and sends the prediction to the SGF.
The measured state variables are directly sent to the SGF.
The SGF output (denoised state estimates) is then used by the RG algorithm to determine the admissible changes in input value.

The results of applying the robust RG algorithm with noisy output measurements is discussed next.
\begin{figure}
    \centering
    \includegraphics[width=\linewidth]{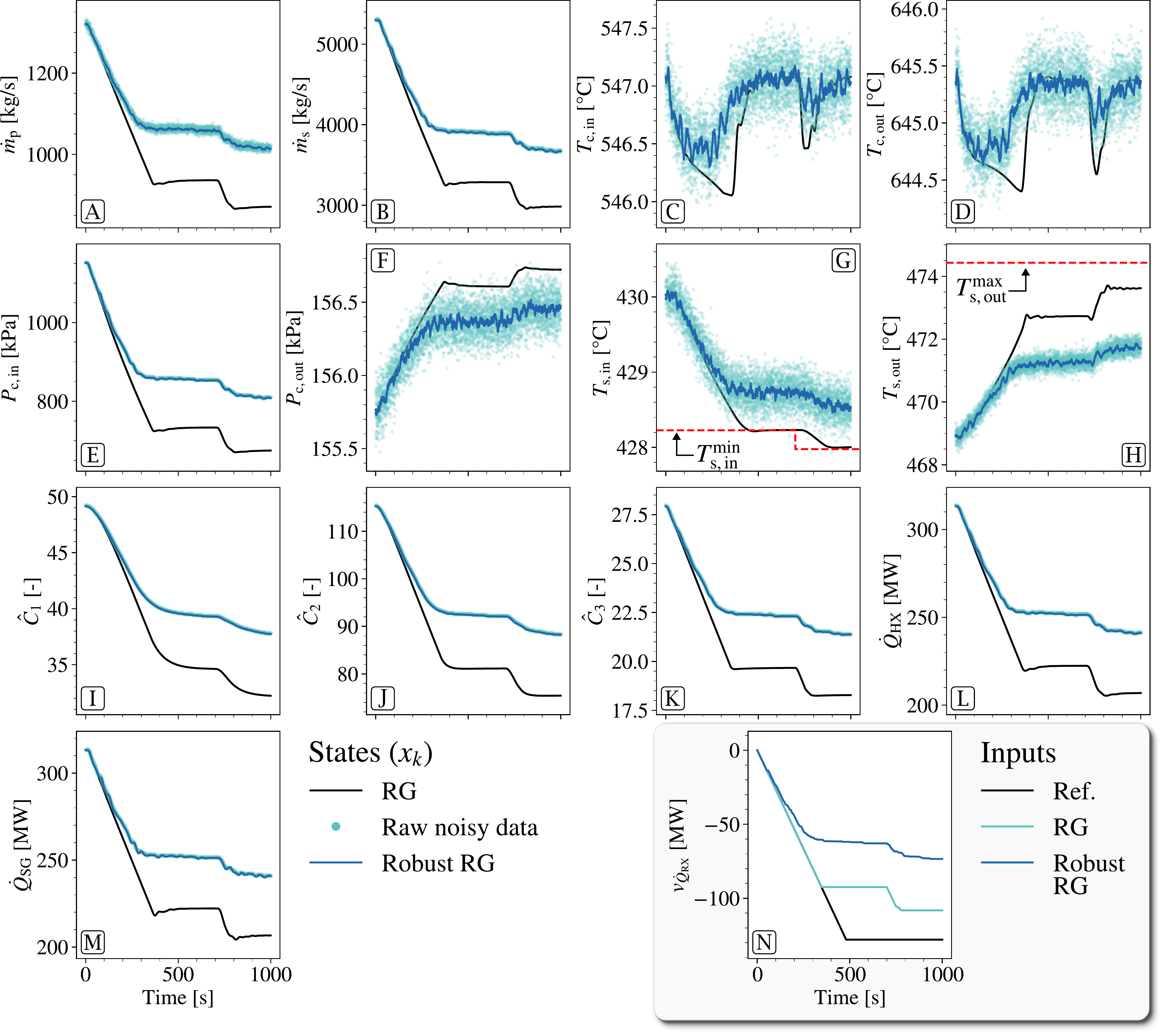}
    \caption{Application of the supervisory control system under the presence of sensor noise.
        Gray box contrasts the reference input value (``Ref.'') against the admitted input value without noise (``RG''), and the admitted input value under sensor noise (``Robust RG'').
        Remaining plots display evolution of state variables for the case without noise (``RG'') and the case with noise (``Robust RG'').
        The unfiltered noisy data is also presented for each state (``Raw noisy data'').
        }
    \label{fig:res-rg-noisy}
\end{figure}
In \cref{fig:res-rg-noisy}, the raw noisy output measurements sent to the UKF and SGF is presented for each $x_k$.
First, the differences in admissible input (\cref{fig:res-rg-noisy}N) is discussed.
The Robust RG input is more conservative than the previous results.
The reason why admissible changes are more conservative is apparent in inspection of the $\Tsi$ plot (\cref{fig:res-rg-noisy}G).
Due to the additional conservatism, even the raw noisy output measurements of $\Tsi$ meet the constraint $\Tsi\leq\Tsimin$.
This implies that if the system designer would want the RG algorithm to be conservative, \ie only trust the sensor measurements, the admissible region will severely contract.
On the other hand, if there is high confidence in the filtered output signals, the admissible region would be larger.
The visualization of the quantitative change in admissible region is discussed in the next section.

\subsection{Quantitative Representation of the Admissible Region}\label{subsec:moas_vis}

\begin{figure}
    \centering
    \includegraphics[width=\linewidth]{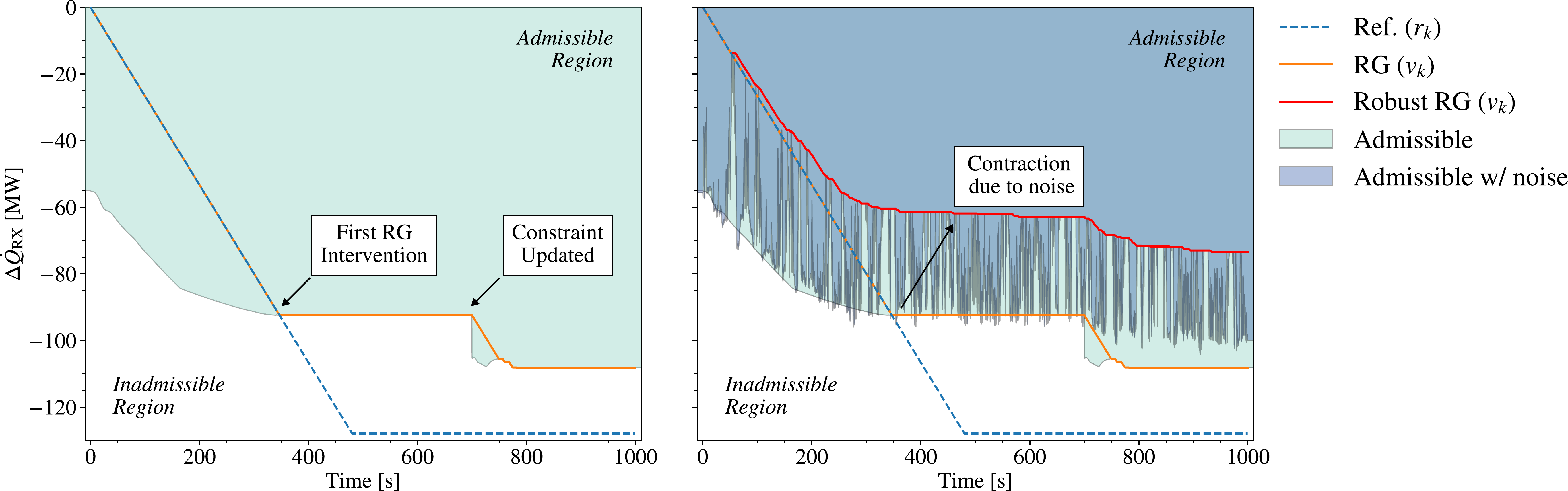}
    \caption{Quantitative evolution of the admissible region ($O_\infty$) during the load-follow transient.
    Left: Evolution during the load-follow transient under noise-free output measurements. 
    The occurrence of the first RG intervention and constraint update is annotated.
    Right: Evolution during the load-follow transient under the presence of output noise. There is a significant contraction in the admissible region as the constraints are applied to the raw (noisy) output signals.
    }
    \label{fig:moas}
\end{figure}

In the Introduction, the problem of assigning a quantitative feasible region during transients was qualitatively visualized in \cref{fig:moas-abs}.
The objective of this work is to develop and demonstrate a supervisory control system that allows constraint enforcement during operational transients for an advanced NPP.
Given the state-space model of the studied system and the set of constraints expressed as linear inequalities, the RG allows quantifying the admissible region and predicting the occurrences of constraint violations, \cref{eq:moas,eq:moas_w}.
In \cref{fig:moas}, the evolution of the admissible region during the 40 \% power drop transient is presented.
The noise-free case is discussed first.
At the beginning of the transient, there is a large margin between the bounds of the admissible region, and the reference setpoint ($r_k$).
In other words, if the supervisory controller allows the requested change in the high-level system input ($v_k=r_k$), no violation of constraint will occur as $t\rightarrow\infty$.
At approximately \SI{347}{\second}, the first RG intervention occurs, annotated in \cref{fig:moas}.
At this point, the reference setpoint intersects the boundary of the admissible region.
The RG algorithm intervenes and does not admit further changes in $v_k$.
As shown in \cref{fig:res-rg}, although this intervention occurs before the $\Tsi\geq\Tsimin$ constraint, as time progresses the equilibrium value of $\Tsi\approx\Tsimin$.
At \SI{700}{\second}, the value of $\Tsimin$ is decreased, and an expansion of the admissible region is realized, allowing further reduction in $v_k$.

The changes in evolution of the admissible region under the presence of output noise is discussed next.
In \cref{fig:moas}, it is apparent that the presence of noise results in a more chaotic evolution of the feasible region.
To quantify the admissible region, $O_\infty$, the current state of the system, $x_k$, is required (\cref{eq:moas_w}).
In the robust RG application, the output measurements are sent to the SGF before the RG block (\cref{fig:RG_ukf}).
Some of the noise that is artificially added to the raw output is still present in the denoised state value (\cref{fig:res-rg-noisy}).
This noise cause the $O_\infty$ boundary itself to be noisy.
Thus, there are two major outcomes from introducing noise in the output measurements.
First, there is a significant contraction in the feasible region as the RG is tasked to enforce $\Tsi\geq\Tsimin+3\sigma_T$ (where $\sigma_T$ is the standard deviation of the Gaussian noise added to temperature measurements, \cref{tab:outpw}).
Second, the evolution of the $O_\infty$ is noisy as further decreases in $v_k$ is only admitted once the noisy state of the system, $x_k$, permits further changes.

\cleardoublepage
\subsection{Advantages \& Drawbacks of Proposed Framework}\label{subsec:adv}

The following are advantages of the proposed framework:
\begin{itemize}
  \item \textbf{Adaptable constraint enforcement}:
  The admissible region is defined from the dynamics of the system and the constraints assigned.
  The RG algorithm allows a point-wise in time enforcement of these constraints.
  Thus, the imposed limits can be updated during operation, either manually by an operator or by a diagnostics module.
  The admissible region will then be updated, and the RG will modify the high-level reference setpoints.
  This capability was demonstrated in \cref{subsec:rg_res} and visualized in \cref{subsec:moas_vis}.
  The ability to enforce time-dependent constraints makes the overall control system adaptable to changing system dynamics and sensor \& actuator performance.
  
  \item \textbf{Noise handling}:
  The RG formulation allows for an explicit assignment of output and process disturbances to quantitatively contract or expand the admissible region.
  This capability was demonstrated in \cref{subsec:res_noise}.
  The ability to account for disturbances is a significant advantage of the RG algorithm.
  Additionally, the anticipated output disturbance that is assigned in \cref{eq:state_w} can also be updated.
  For example, if a diagnostics module detects that a thermocouple has degraded, the admissible region would be updated to account for the additional uncertainty in measurements.
  
  \item \textbf{Hierarchical control}:
  The proposed architecture is constituted by a supervisory layer and a low-level controller layer (\cref{fig:RG}).
  The former is tasked with admitting changes to high-level reference setpoints, whereas the latter has tracking and regulation duties.
  This separation allows the low-level controllers to be tuned and optimized independently.
  In the event that manual operation is desired, the supervisory layer can be bypassed.
  This is a significant benefit over end-to-end control paradigms, \eg classic MPC, where the high-level inputs and low-level actuation are handled by a single algorithm.

  \item \textbf{Deployment of Autonomous Operation}:
  As described by \citet{Wood2017}, Automated control involves self-action, while Autonomous control involves independent action, greater fault management, more embedded planning, and even self-healing or adaptation.
  The first supervisory control systems \citep{demore1977plant,White1989} were meant to enhance the level of automation of existing architectures.
  Their objective was to synchronize the different control modules, \ie “reactor control”, “primary flow control” and “turbine and bypass control”, during normal operation and complicated sequences (\eg reactor startup).
  The supervisory control system proposed in this work allows (1) enforcing time-dependent constraints on process variables, (2) separating tracking and constraint-enforcement, and (3) optimizing the Multi Input Multi Output control problem.
  These features are propaedeutic to the deployment of an Autonomous Operation framework since external decision-making or diagnostics modules can be directly interfaced with the control module to update constraints and the ensuing admissible region during operation, as shown in \cref{fig:moas}.

\end{itemize}
Additionally, the following are drawbacks of the proposed control system:
\begin{itemize}
  \item \textbf{Static system matrices}:
  Currently, the matrices identified by the DMDc algorithm are used by the RG algorithm via \cref{eq:state,eq:moas}.
  The matrices are not updated during the load-follow transient.
  This modeling choice implies that a \textit{constant} linear approximation is used to describe the dynamics of the gFHR.
  This is valid on short time-scales, but it might be inaccurate for larger time-scales (weeks to months) as the dynamics of a physical plant might change, \eg due to fuel burnup or accumulation of fouling.
  A dedicated procedure for updating the state-space matrices must be incorporated.
\end{itemize}

\section{Concluding Remarks}\label{sec:conclusions}

To make advanced reactors more competitive in deregulated power markets characterized by a significant presence of intermittent energy sources, flexible operation is sought.
Autonomous control during routine power transients will relieve plant operators from repetitive, error-prone task and might help them determining the optimal course of action. 
In this work, the operation of a \SI{320}{\mega\watt} Pebble-bed Fluoride-cooled High-temperature reactor is studied. 
The reactor is modeled using the best-estimate system-code SAM.
During the load-follow transient, several process variables are carefully controlled to ensure thermal hydraulic and neutronic conditions in the primary circuit.
However, there are several uncontrolled process variables that deviate significantly from their steady-state 100 \% load values.

In this work, a supervisory control system to constrain load changes that might cause excessive variations of uncontrolled process variables is designed.
To achieve this objective, several algorithms were integrated.
The RG algorithm was used as the supervisory layer to quantify the admissible region given a set of constraints, and accept or reject changes in high-level inputs.
The DMDc algorithm was used as a data-driven approach to system identification.
The SFFS algorithm was used to identify the appropriate set of process variables to be modeled by the DMDc algorithm.
The UKF algorithm was used to infer unobserved DNP values using the non-linear 6-group PKE model.
The SGF was used to denoise signals.
The integration of these algorithms successfully realized the abstract objective in \cref{fig:moas-abs} to a quantitative result in \cref{fig:moas}.
The supervisory control system was able to intervene during the 40\% power drop to avoid violation of temperature constraints assigned to the secondary loop.
The constraints were also updated during the transient to demonstrate adaptability.
Under the presence of sensor noise, the robustness of the control system was demonstrated as well.
Advantages of the proposed control system are summarized in \cref{subsec:adv}.

There are other tasks towards achieving autonomous operation for advanced reactors.
There are auxiliary tasks that can be incorporated into the overall autonomous operation framework.
For example, diagnosing the degradation of an actuator (\eg pumps or valves), component (\eg heat exchanger) performance degradation, or sensor drift, can inform the supervisory layer on constraints that need to be updated or added.
Additionally, the application of the supervisory control system to a physical system is necessary to assess its capabilities.
A large concern would have been the impact of sensor noise, which has been mitigated with the demonstration of the robust RG.
Finally, as elaborated in \cref{subsec:adv}, a procedure to update the state-space representation over long time-scales, must be explored. 

\section*{CRediT authorship contribution statement}
\textbf{Akshay J.~Dave}: Methodology, Software, Investigation, Writing - Original Draft, Visualization. 
\textbf{Taeseung Lee:} Methodology for Control Strategy, Writing - Review \& Editing.
\textbf{Roberto Ponciroli:} Methodology, Conceptualization, Funding acquisition, Supervision, Writing - Review \& Editing.
\textbf{Richard B.~Vilim:} Conceptualization, Funding acquisition, Supervision, Project administration.

\section*{Acknowledgements}
This work is supported by US Department of Energy Advanced Research Projects Agency–Energy Project 2174-1556, and US Department of Energy Nuclear Energy Enabling Technologies Project 20-19321.

\bibliographystyle{unsrtnat}
\bibliography{references}

\end{document}